\documentclass[seceqn,dvips]{arxstspdf}
\usepackage{flushend}
\usepackage{stfloats}

\volume{25}
\issue{4}
\pubyear{2010}
\firstpage{506}
\lastpage{516}
\doi{10.1214/10-STS341}

\begin{document}
\begin{frontmatter}

\title{From EM to Data Augmentation: The~Emergence of MCMC Bayesian
Computation in the 1980s}
\runtitle{Emergence of MCMC Bayesian
Computation}

\begin{aug}
\author{\fnms{Martin A.} \snm{Tanner}\corref{}\ead[label=e1]{mat132@northwestern.edu}}
\and
\author{\fnms{Wing H.} \snm{Wong}\ead[label=e2]{whwong@stanford.edu}}
\runauthor{M. A. Tanner and W. H. Wong}

\affiliation{Northwestern University and Stanford University}

\address{Martin A. Tanner is Professor,
Department of Statistics, Northwestern University, Evanston, Illinois 60208, USA
\printead{e1}.
Wing H. Wong is Professor,
Departments of Statistics and Health Research and Policy, Stanford University,
Stanford, California 94305, USA
\printead{e2}.}

\end{aug}

%
\begin{abstract}
It was known from Metropolis et al. [\textit{J. Chem. Phys.}
\textbf{21} (1953) 1087--1092] that one can sample from a distribution by
performing Monte Carlo simulation from a Markov chain whose equilibrium
distribution is equal to the target distribution. However, it took
several decades before the statistical community embraced Markov chain
Monte Carlo (MCMC) as a general computational tool in Bayesian
inference. The usual reasons that are advanced to explain why
statisticians were slow to catch on to the method include lack of
computing power and unfamiliarity with the early dynamic Monte Carlo
papers in the statistical physics literature. We argue that there was a
deeper reason, namely, that the structure of problems in the
statistical mechanics and those in the standard statistical literature
are different. To make the methods usable in standard Bayesian
problems, one had to exploit the power that comes from the introduction
of judiciously chosen auxiliary variables and collective moves. This
paper examines the development in the critical period 1980--1990, when
the ideas of Markov chain simulation from the statistical physics
literature and the latent variable formulation in maximum likelihood
computation (i.e., EM algorithm) came together to spark the widespread
application of MCMC methods in Bayesian computation.
\end{abstract}

%

\end{frontmatter}

\section{Introduction}

This paper surveys the historical development of MCMC methodology
during a key time period in Bayesian computation. As of the mid-1980s,
the Baye\-sian community was focused on Gaussian quadrature type methods,
Laplace approximations and variants of importance sampling as the main
computational tools in Bayesian analysis. Among more dogmatic
Bayesians, the use of Monte Carlo was met with resistance and viewed as
antithetical to Bayesian principles. MCMC techniques published in the
statistical physics and image analysis literature were seen by the
Bayesian computational community as techniques for specialized
problems. However, by the early 1990s MCMC-based approaches have become
a mainstay in computational Bayesian inference. The purpose of this
paper is to review the events that led to this remarkable development.
In particular, we examine the critical decade of 1980--1990 when the
ideas of Markov chain simulation from the statistical physics
literature and the latent variable formulation in maximum likelihood
computation (i.e., EM algorithm) came together to spark the widespread
application of MCMC methods in Bayesian computation.

\section{Some Pre-History}

\subsection{Markov Chain Monte Carlo}

The origin of MCMC can be traced to the early 1950s when physicists
were faced with the need to numerically study the properties of many
particle systems. The state of the system is represented by a vector
$x$ = $( x_1, x_2 , \ldots, x_n )$, where $x_i$ is the coordinate of the
ith particle in the system and the goal is to study properties such as
pressure and kinetic energy, which can be obtained from computation of
the averaged values of suitably defined functions of the state vector.
The averaging is weighted with respect to the canonical weight
$\exp(-E(x)/kT)$, where the constants $k$ and $T$ denote the Boltzmann
constant and the temperature, respectively. The physics of the system
is encoded in the form of the energy function. For example, in a simple
liquid model one has the energy $E(x)$ = $(1/2) \sum\sum_ {i \neq j}
V(| x_i - x_j |)$, where $V(\cdot)$ is a potential function giving the
dependence of pair-wise interaction energy on the distance between two
particles. Metropolis et al. (\citeyear{metropolisetal}) introduce the first Markov chain
Monte Carlo method in this context by making sequential moves of the
state vector by changing one particle at a time. In each move, a random
change of a particle is proposed, say, by changing to a position chosen
within a fixed distance from its current position, and the proposed
change is either accepted or rejected according to a randomized
decision that depends on how much the energy of the system is changed
by such a move. Metropolis et al. justified the method via the concepts
of ergodicity and detailed balance as in kinetic theory. Although they
did not explicitly mention ``Markov chain,'' it is easy to translate
their formulation to the terminology of modern Markov chain theory. In
subsequent development, this method was applied to a variety of
physical systems such as magnetic spins, polymers, molecular fluids and
various condense matter systems (reviewed in Binder, \citeyear{binder}), but all
these applications share the characteristics that $n$ is
large\footnote{In the words of Geman and Geman (\citeyear{gemangeman}), ``The Metropolis
algorithm and others like it were invented to study the equilibrium
properties, especially ensemble averages, time-evolution and
low-temperature behavior, of very large systems of essentially
identical, interacting components, such as molecules in a gas or atoms
in binary alloys.''} and the $n$ components are homogeneous in the
sense that each takes value in the same space (say, 6-dimensional phase
space, or up/down spin space, etc.) and interacts in identical manner
with other components according to the same physical law as specified
by the energy function. These characteristics made it difficult to
recognize how the method can be of use in a typical Bayesian
statistical inference problem where the form of the posterior
distribution is very different from the Boltzmann distributions arising
from physics. For this reason, although the probability and
statistical community was aware of MCMC very early on (Hammersley and
Handscomb, \citeyear{hammersleyhandscomb}\footnote{Interestingly, Hammersley and Handscomb
(\citeyear{hammersleyhandscomb})
present the discrete version of the Metropolis algorithm in the chapter
entitled ``Problems in statistical mechanics,'' rather than in the
chapter on ``Principles of the Monte Carlo method'' which covers topics
such as crude Monte Carlo, stratified sampling, importance sampling,
control variates, antithetic variates and regression methods.
Similarly, in another popular textbook, Rubinstein (\citeyear{rubinstein}) presents the
discrete version of the Metropolis algorithm toward the end of his
book, embedded in an algorithm for `global' optimization, rather than
earlier in, for example, Chapter 3, ``Random variate generation'' or
in Chapter 4, ``Monte Carlo integration and variance reduction
techniques.'' Also, Ripley (\citeyear{ripley87}) presents the discrete version of the
Metropolis algorithm in the section ``Metropolis' method and random
fields'' (but not in the previous chapter on simulating random
variables or in the following chapter on Monte Carlo integration and
importance sampling).}) and had in fact made key contributions
to its theoretical development (Hastings,~\citeyear{hastings}), the method was not
applied\footnote{Several authors have argued that this delay can be
attributed to ``lack of appropriate computing power'' (quoting Robert
and Casella, \citeyear{robertcasella10}). Kass (\citeyear{kass}) in his \textit{JASA} review
of the book
\textit{Markov Chain Monte Carlo in Practice} remarks on page 1645: ``I
believe that MCMC became important in statistics when it did for the
simple reason that in the early 1990s large numbers of researchers
could implement it on their desktops for interesting, nontrivial
problems\ldots\ I would suggest that the timing of their growth in
popularity is explained primarily by computing technology.'' [This
argument is quoted almost verbatim by Hitchcock (\citeyear{hitchcock})---see Gubernatis
(\citeyear{gubernatis}) who presents several ``first-hand'' accounts of the history of
the Metropolis algorithm, as well as discusses why the algorithm
received scant use in the 10--15 years following its
development.]\

This response does not explain why in 1987, when research universities
had long since transitioned away from the university mainframe (which
billed for every second of CPU usage, every page of memory, etc.) to
departmental workstations (typically VAX 780's or 750's, which were
essentially free to faculty), the key computational Bayesian methods
advocated at the time were based on quadrature or Laplace
approximations or importance sampling (see the \hyperref[appendix]{Appendix}). Of course, one
could not carry a VAX 750 around in a briefcase and it was not as
powerful as current PCs, but one could log on from home or from the
office and submit a job to run overnight (or over a weekend) without
having to deal with a charge-back system. Clearly, the state of
computing in the early 1980s did not detract from the work of Geman and
Geman (\citeyear{gemangeman}).

It may not have been a lack of computing power \textit{per se} that
contributed to this developmental delay, but possibly a concern
regarding optimal algorithms. In this regard, Geman (1988a) notes:
``\ldots\ we view Monte Carlo optimization techniques as \textit{research
tools}. They are poor substitutes for the efficient dedicated
algorithms that should be developed when facing applications involving
a flow of data and a need for speedy analysis.'' Efron (\citeyear{efron}) mentions
that Tukey wanted to call the \textit{bootstrap} the \textit{shotgun},
because it ``can blow off the head of any problem if the statistician
can stand the resulting mess.'' The same can be said for
Metropolis--Hastings algorithms, as they can solve any simulation
problem, but as noted by Geman (\citeyear{geman88a}), there is a strong sense that
one could find a more efficient approach to implementing the simulation
\textit{in the specific case}.

Robert and Casella (\citeyear{robertcasella10}) also cite ``lack of background on Markov
chains.'' We question this as well, for by the mid-1970s it was
standard practice to require a stochastic process course based on such
texts as Karlin and Taylor (\citeyear{karlintaylor}) as part of the degree requirements
for a Ph.D. in Statistics. The treatment in Karlin and Taylor (\citeyear{karlintaylor}), for
example, provides a more than sufficient background to understand the
related material in Hammersley and Handscomb (\citeyear{hammersleyhandscomb}).} to Bayesian
inference until the 1980s.

\subsection{Latent Variable Methods in Likelihood Inference: The EM Algorithm}

During the 1960s and 1970s, statisticians developed an approach to
maximum likelihood computation that is quite effective in many popular
statistical models. The approach was based on the introduction of
latent variables into the problem so as to make it feasible to compute
the MLE if the latent variable value were known. Equivalently, if one
regards the latent variable as ``missing data'', then this approach
relies on the simplicity of inference based on the ``complete data'' to
design an iterative algorithm to compute the maximum likelihood
estimate and the associated standard errors. This development
culminated in the publication of the extremely influential paper
Dempster, Laird and Rubin (\citeyear{dempsteretal}). A review of earlier research treating
specific examples was presented in that paper, as well as the
associated discussion. The high impact of Dempster, Laird and Rubin stems from
its compelling demonstration that a wide variety of seemingly unrelated
problems in standard statistical inference, including multinomial
sampling, normal linear models with missing values, grouping and
truncation, mixture problems and hierarchical models, can all be
encompassed within this latent variable framework and thus become
computationally feasible using the same algorithm (called the EM
algorithm by Dempster, Laird and Rubin) for MLE inference.

Because of its influence in later MCMC methods on the same set of
problems, we briefly review a simplified formulation of the EM
algorithm: Let $y$ be the observed data vector, $p _{ \theta} (y)$ be
the density of $y$, and we are interested in the inference regarding
$\theta$. Two conditions are assumed for the application of the EM.
First, it is assumed that although the likelihood $L( \theta|y) = p
_ {\theta} (y) $ may be hard to work with, one can introduce a latent
(i.e., unobserved) variable $z$ so that the likelihood $ L( \theta
|y,z) = p _ {\theta} (y, z)$ based on the value of $y$ and $z$
becomes easy to optimize as a function of $\theta$. In fact, for
simplicity, we assume that $p _ {\theta} (y, z)$ is an exponential
family distribution. The second condition is that for any fixed
parameter value $\theta$, it is possible to compute the expectation of
the sufficient statistics of the exponential family, where the
expectation is over $z$ under the assumption that $z$ is distributed
according to its conditional distribution $p _ \theta(z|y)$. We will
see below that these conditions are closely related to the ones under
which the most popular form of MCMC algorithm for Bayesian computation,
namely, the Gibbs sampler, is applicable.

\section{Emergence of MCMC in Bayesian Computation in the 1980s}

\subsection{The State of Bayesian Computation in~the~Mid-1980s}

The early 1980s was an active period in the development of Bayesian
computational methods. In addition to the traditional approach that
relied on the use of conjugate priors to obtain analytically tractable
posterior distributions, significant progress was made in numerical
approximations to the posterior distribution. We now briefly review the
main approaches.

In many problems it is easy to evaluate the joint posterior density up
to a constant of proportionality. The difficulty is to obtain posterior
moments and marginal distributions of selected parameters of interest.
Numerical integration methods were developed to obtain these quantities
from the joint posterior. In particular, Naylor and Smith (\citeyear{naylorsmith}) and
Smith et al. (\citeyear{smithetal85}) advocated the use of Gaussian quadrature which
would be the correct choice in large sample situations when the
posterior is approximately normal. Alternatively, Kloek and van Dijk
(\citeyear{kloek1}, \citeyear{kloek2}) proposed the use of importance sampling to carry out the
integration, and applied the method systematically in the context of
simultaneous equation models. Many novel variations were experimented
with in both approaches, including the important idea of adaptation
where a preliminary integration was used to guide the choice of
parameters (grid points, importance function, etc.) in a second round
of integration.

Another influential work was Tierney and Kadane (\citeyear{tierneykadane}) which uses the
technique of Laplace approximation to obtain accurate approximations
for posterior moments and marginal densities; albeit in contrast to the
other approaches, the accuracy of this approximation is determined by
the sample size and not under the control of the Bayesian analyst.

These efforts demonstrated that accurate numerical approximation to
marginal inference can be obtained in problems with moderate
dimensional parameter space [e.g., Smith et al. (\citeyear{smithetal85}) report success
on problems with up to 6 dimensions] and created a great deal of
excitement in the prospect of computational Bayesian inference (see,
e.g., Zellner, \citeyear{zellner}). On the other hand, a~review of the
writings of leading Bayesian statisticians in this period reveal no
awareness of the promise of the MCMC approach that would soon emerge as
a dominate tool in Bayesian computation. In fact, among more dogmatic
Bayesians,\footnote{ For a more detailed review of the prevailing
approaches and views on Bayesian computation through the late 1980s,
see the \hyperref[appendix]{Appendix}.} the use of Monte Carlo was met with resistance and
viewed as antithetical to Bayesian principles.

We are now ready to discuss the specific developments that sparked the
emergence of MCMC methodology in statistics.

\subsection{Formulation of the Gibbs Sampler}

In 1984, Geman and Geman published a paper on the topic of Bayesian
image analysis (Geman and Geman,~\citeyear{gemangeman}). Beyond its immediate and large
impact in image analysis, this paper is significant for several results
of more general interest, including a proof of the convergence of
simulated annealing, and the introduction of the Gibbs sampler.

We briefly review how the Gibbs sampler emerged in this context. The
authors began by drawing an analogy between images and statistical
mechanics systems. Pixel gray levels and edge elements were regarded as
random variables and an energy function based on local characteristics
of the image was used to represent prior information on the image such
as piece-wise smoothness. Because interaction energy terms involved
only local neighbors, the conditional distribution of a variable given
the remaining components of the image depends only on its local
neighbors, and is therefore easy to sample from. Such a distribution,
for the systems of image pixels, is similar to the canonical
distribution in statistical mechanics studied by Boltzmann and Gibbs,
and it is thus called a Gibbs distribution for the image.

Next, the authors analyzed the statistical problem of how to restore
the image from an observed image which is a degradation of true image
through the processes of local blurring and noise contamination. They
showed that the posterior distribution of true image given the observed
data is also a Gibbs distribution whose energy function still involves
only local interactions. Geman and Geman proposed to generate images
from this posterior distribution by iteratively sampling each image
element from its conditional distribution given the rest of the image,
which is easy to do because the distribution is still Gibbs. They call
this iterative conditional sampling algorithm the Gibbs sampler. For
the history of Bayesian computation, this was a pivotal step---although
similar algorithms were already in use in the physics literature; to
our knowledge, this work represented the first proposal to use MCMC to
simulate from a posterior distribution. On the other hand, because the
Gibbs model for images is so similar to the (highly specialized)
statistical physics models, it was not apparent that this approach
could be effective in traditional statistical models (see the \hyperref[appendix]{Appendix}).

\subsection{Introduction of Latent Variables and Collective Moves}

The use of iterative sampling for Bayesian inference in traditional
statistical models was first demonstrated in Tanner and Wong (\citeyear{tannerwong}).
The problems treated in this work, such as normal covariance estimation
with missing data, latent class models, etc., were of the type familiar
to mainstream statisticians of the time. A characteristic of many of
these problems was that the likelihood is hard to compute (thus not
amenable to MCMC directly). To perform Bayesian analysis on these
models, the authors embedded them in the setting of the EM algorithm
where a latent variable $z$ can be introduced to simplify the inference
of the parameter $\theta$. They started from the equations
\begin{eqnarray}\label{e31}
p(\theta|y) &=& \int p( \theta|y,z) p(z|y)\,dz,\\\label{e32}
p(z|y) &=& \int p(z|\theta, y)p( \theta|y)\,d\theta.
\end{eqnarray}

Recall that the conditions needed for the EM to work well are that $p
_{\theta}(y, z)$ is simple to work with as a function of $\theta$, and
$p _ {\theta}(z|y)$ is easy to work with as a function of $z$. The
first condition usually implies that the complete data posterior $p(
\theta|y,z)$ is also easy to work with. Thus, (\ref{e31}) can be
approximated as a mixture of $p( \theta|y,z)$ over a set of values
(mixture values) for~$z$ drawn from (\ref{e32}). Similarly, (\ref{e32}) is
approximated as a mixture of $p(z| \theta, y)$ over mixture values for
$\theta$ drawn from (\ref{e31}). This led the authors to propose an iterated
sampling scheme to construct approximations to $p( \theta|y)$ and
$p(z|y)$ simultaneously. In each step of the iteration, one draws a
sample of values with replacement from the mixture values for $z$ (or
$\theta$), and then conditional on each such $z$, draws $\theta$ (or
$z$) from $p( \theta|y,z)$ [or $p(z| \theta, y)$].

This computation is almost identical to a version of the Gibbs sampler
that iterates between the sampling of $p( \theta|y,z)$ and $p(z|
\theta, y)$. In fact, if the sampling from the mixture values for $z$
(or $ \theta$) was done without replacement rather than with
replacement, as suggested by Morris (\citeyear{morris87}), then one would have exactly
a population of independently run Gibbs samplers. The authors also
noted\footnote{Quoting Tanner and Wong (\citeyear{tannerwong}): ``To see this, consider
the extreme case in which $m = 1$, so that iteration $i$
produces only one value $ \theta(i)$. In this case, $\theta(i)\,
(i = 1, 2, \ldots)$ forms a Markov process with transition
function equal to $K( \theta, \phi)$, as defined in (2.3). Under
the regularity conditions of Section 6, this is an ergodic Markov
process with an equilibrium distribution satisfying the fixed point
equation given in (2.3).'' Equation (2.3) presents the Markov
transition function ${K( \theta, \phi)}$ (see also their Markov
transition operator discussion in Remark 4) via the expression
$g(\theta) =  \int K( \theta, \phi)g( \phi)\,d \phi$, where
${K( \theta, \phi)} = \int {p( \theta | z,y) p(z | \phi
,y)\,dz}$.} the connection to the equilibrium distribution of a Markov
chain, but they did not employ it as the main mathematical framework in
their analysis. In any case, a prominent aspect of its relevance lies
in the explicit introduction of the latent variable $z$, which may or
may not be part of the data vector or the parameter vector of the
original statistical model, to create an iterative sampling scheme for
the Bayesian inference of the original parameter $\theta$. Tanner and
Wong referred to this aspect of the design of the algorithm as ``data
augmentation.'' A judicious choice of latent variables can allow one to
sample from the posterior $p( \theta|y)$ in cases where direct MCMC
methods, including the Gibbs sampler, may not even be applicable
because of difficulty in evaluating $p( \theta|y)$.

As a discussant of Tanner and Wong (\citeyear{tannerwong}), Morris~(\citeyear{morris87}) makes several
key observations of great relevance to MCMC Bayesian computing. In
addition to suggesting a version of data augmentation that is the same
as parallel Gibbs sampling, he emphasizes that (just as in the EM
context) the augmentation is not limited to missing data, but can be
done with parameters as well: ``and to emphasize that their `missing
data' concept can be used to include unknown parameters or latent
data.'' As an illustration of the data augmentation algorithm, Morris
(\citeyear{morris87}) presents what we would now call the Gibbs sampler for a
three-stage hierarchical model with $k + 1$ parameters. At the first
level of his model, $y_i | \theta_i $ are distributed independently as
$N ( \theta_i , V_i )$, for $i = 1, \ldots, k$ ($V_i$ known). At the
second stage, $ \theta_i | A $ are i.i.d. $N(0,A), i = 1,\ldots, k$.
At the final stage, $A$ is distributed as a completely known
distribution. Morris then says, ``Let initial values $A^{(1)},\ldots,
A^{(m)}$ be given. The posterior distribution of $\theta$ given
$(y,A)$ is normally distributed and the P step samples $ \theta_i^{(j)}
\sim N((1-B_i^{(j)} )y_i , V_i (1-B_i^{(j)}))$ for $i = 1,
\ldots, k$, $j = 1, \ldots,\break m$, independently, with ${B_i^{(j)}} = V_i
/ (V_i + A^{(j)} )\ldots.$'' For the $A$ parameter he writes, ``The I
step (1.5), therefore, samples new values $A^{(1)},\ldots, A^{(m)}$
according to $A^{(j)}\sim (\lambda + \Vert \theta^{(j)} \Vert^2)/
{\chi_{k+q} ^2} $ for $j = 1, \ldots, m$, with $ {\chi_{k+q} ^2} $
sampled independently for each $j$, $\Vert\theta\Vert^2$ denoting the sum
of squares.''

Although no new theory beyond MCMC is needed for the analysis of a
sampling algorithm designed to include latent variables, the nature of
the resulting process may be drastically different from the traditional
MCMC processes, even in cases when the posterior $p ( \theta| y )$ is
computable and therefore amenable to direct MCMC analysis. Consider,
for example, a linear model $y = x \beta + \varepsilon$, where the
errors are independently distributed according to Student's $t$ with a
fixed degree of freedom. In this case the joint posterior density for
$\beta$ is computable and one can apply the standard Metropolis sampler
to sample from it, by iteratively proposing to change the $\beta$
vector, one component at a time. However, the sampler may be easily
trapped at local maxima. What is worse, the moves may be exceedingly
localized (therefore slow) if there is serious collinearity. On the
other hand, by regarding the error as a gamma-mixture of normal
variables, one can condition on the gamma variables and generate the
whole vector $\beta$ (i.e., a collective move), which allows large
moves even in the presence of collinearity.\footnote{For simplicity, the
errors are unscaled in our example. If the errors have a scaling
parameter, then it will have correlation with the gamma variables and
this will slow down the convergence. More advanced augmentation schemes
can be created to break this correlation and accelerate convergence;
see Liu, Rubin and Wu (\citeyear{chanhai}), Liu and Wu (\citeyear{liu}) and van Dyk and Meng
(\citeyear{vandyk}).}

Interestingly, at about the same time, Swendsen and Wang (\citeyear{swendsenwang}) also
introduced the use of latent variables (called auxiliary variables) in
the setting of a statistical mechanics system. This work deals with the
Potts model of spins on a lattice. The authors introduced bond
variables between spins and then alternated between the sampling of the
two types of variables, spins and bonds. By conditioning on the bonds,
they were able to make more global changes of the spin configuration by
simultaneously updating a whole cluster of spins that are connected by
active bonds (i.e., a collective move). In this way they were able to
dramatically reduce the correlation time of the resulting Markov
process for simulating a two-dimensional Ising model. Justifiably, this
work is widely regarded as a breakthrough in dynamic Monte Carlo
methods in statistical physics.

\subsection{A Synthesis}

Above we described how MCMC Bayesian computation arose in the 1980s
from two independent sources, the statistical physics heritage as
represented by Geman and Geman (\citeyear{gemangeman}), and the EM heritage as
represented by Tanner and Wong (\citeyear{tannerwong}). A synthesis of these two
traditions occurred in the important work of Gelfand and Smith (\citeyear{gelfandsmith}).
Like the former, they employed the Gibbs sampling version of MCMC. Like
the latter, they focused on traditional statistical models and relied
on the use of latent variables to create iterative sampling schemes.
Their paper\footnote{Also of note are the papers of Li (\citeyear{li8}) who used
a multi-component Gibbs sampler to perform multiple imputation from the
posterior, Gelman and King (\citeyear{gelmanking}) who employed MCMC to analyze
hierarchical models of voting data across districts and Spiegelhalter
and Lauritzen (\citeyear{spiegellaur}) who treated graphical models. The latter appeared
to have inspired the popular MCMC software
\texttt{BUGS} (Lunn et al., \citeyear{lunnetal}).} provided many examples to
illustrate the ease of use\footnote{Smith (\citeyear{smith91}) notes that regarding
the Gibbs sampler: ``Substantial iterative computation is then
required, but the need for sophisticated numerical understanding on the
part of the statistical analyst is obviated.'' Gelfand et al.
(\citeyear{gelfandetal})
also comment that sampling methods have a ``hidden'' efficiency: ``In
addition, in the case of most of the more sophisticated techniques,
substantial fresh effort is required (including, in some cases,
beginning the analysis anew) if the focus of inferential interest
changes\ldots.'' Tanner and Wong~(\citeyear{tannerwong}, pages 533 and 549), in their
normal variance--covariance matrix example, make the same point when
they note the ease in which one may examine the posterior distribution
of any function of the variance--covariance matrix, such as the smallest
eigenvalue (see their Figure 1 in the Rejoinder).} and effectiveness of
iterative sampling, and clarified the relation between the data
augmentation algorithm and the Gibbs sampler.

The framing of data augmentation as MCMC also raised some new and
interesting theoretical issues in the analysis of the MCMC output. For
example, it follows from (\ref{e32}) that in data augmentation the estimate
of an expectation $E(g( \theta)|y)$ is given by $\frac{1}{n}
\sum_{i=1}^{n} E(g( \theta)|y, z_i )$, where the $z_i$'s are the
currently sampled values for the latent variable $z$. Gelfand and Smith
refer to the use of this estimate, instead of the usual estimate
$\frac{1}{n} \sum_{i=1}^{n} g( \theta_i)$, as Rao--Blackwellization.
They reasoned that if the $z_i$'s are independently drawn, as in a
final iteration of the data augmentation algorithm, then clearly
Rao--Blackwellization will reduce estimation error. They did not
analyze the situation when the samples are dependent, as when the
samples are generated from the Gibbs sampling process. The superiority
of the Rao--Blackwelli\-zed estimator in the two-component Gibbs sampler
was later established in Liu, Wong and Kong (\citeyear{liuetal})---see also
Geyer (\citeyear{geyer95}).

After the publication of Gelfand and Smith's influential paper, many
mainstream statisticians began to adapt the use of MCMC in their own
research, and the results in these early applications quickly
established MCMC as a dominant methodology in Bayesian computation.
However, it should be noted that in any given problem there could be a
great many ways to formulate a MCMC sampler. In simulating an Ising
model, for example, one can try to flip each spin conditional on the
rest, or flip a whole set of spins connected by (artificially
introduced) bonds that are sampled alternatively with the spins. The
effectiveness of the Swendsen and Wang (\citeyear{swendsenwang}) algorithm in the Ising
model does not simply stem from the fact that it is a Gibbs sampler,
but rather depends critically on the clever design of the specific form
of the sampler. Likewise, a~large part of the success of MCMC in the
early 1990s was based on versions of Gibbs samplers that were designed
to exploit the special structure of statistical problems in the style
of the EM and data augmentation algorithms. Thus, the emergence of MCMC
in mainstream Bayesian inference has depended as much on the
introduction of the mathematically elegant MCMC formalism as on the
realization that the structure of many common statistical models can be
fruitfully exploited to design versions of the algorithm that are
feasible and effective for these models.

The appearance of Gelfand and Smith (\citeyear{gelfandsmith}) mar\-ked the end of the
emergence of the MCMC approach to the study of posterior distributions,
and the beginning of an exciting period, lasting to this day, of the
application of this approach to a vast array of problems, including
inference in non-parametric problems. Advanced techniques have also
been developed in this framework to accelerate convergence.
Statisticians, no longer laggards in MCMC methodology, now rival
physicists in the advancement of MCMC methodology.\footnote{For example,
Geyer (\citeyear{geyer91}) introduced parallel tempering ahead of related concepts in
the physics literature---see Hukushima and Nemoto (\citeyear{huku}).} It is our
hope that this paper will serve as a useful historical context to
understand current developments.

\appendix
\section*{Appendix}\label{appendix}

In this appendix we present three key resources that define the
state-of-the-art Bayesian computing as of the mid to late 1980s.
\setcounter{subsection}{0}
\setattribute{citeyear}{font}{\sffamily}
\subsection{\texorpdfstring{Smith et al.(\protect\citeyear{smithetal85})}{Smith et al. (1985)}}
\setattribute{citeyear}{font}{\relax}

A key reference that catalogs the tools in the Baye\-sian's armamentarium
as of 1985 is the paper by Adrian F. M. Smith and colleagues entitled
``The implementation of the Bayesian paradigm'' (see Smith et al.,
\citeyear{smithetal85}). After providing an overview of the goals of Bayesian
computation, the authors critically review a number of implementation
strategies: (1) exact analytic implementation based on conjugate
priors; (2) large sample normal approximation; (3) alternative
asymptotic approximations including discussion of a preprint of Tierney
and Kadane (\citeyear{tierneykadane}) outlining the Laplace approach; (4) Monte Carlo
integration---specifically importance sampling; (5) quadrature
methodology---specifically the approach of Naylor and Smith (\citeyear{naylorsmith}) based
on the Gauss--Hermite product rules; and (6)~successive transformations
of the parameters of the model to achieve normality. The remainder of
Smith et al.~(\citeyear{smithetal85}) reviews in considerable detail the Naylor and Smith
(\citeyear{naylorsmith}) strategy based on Gaussian quadrature, as well as presents
several interesting examples. The authors conclude with the following
remark: ``novel numerical integration techniques together with
efficient graphical procedures are now making Bayesian analysis
practical for a wide range of problems.''

\subsection{The 1987 Special Issue of \textit{JRSS D} (\textit{The~Statistician})}

In 1987, the \textit{Journal of the Royal Statistical Society Series D
(The Statistician)} published a special issue entitled ``Practical
Bayesian statistics,'' edited by Gopal K. Kanji with technical editors
Adrian F. M. Smith and A. P. Dawid. Following up on the report of
Smith et al. (\citeyear{smithetal85}), Smith et al. (\citeyear{smithetal87}) discuss an adaptive approach
where Gauss--Hermite quadrature methods are combined with parameter
transformations in an iterative manner, namely, successive
transformations are determined by the estimated variance--covariance
matrix of the previous iteration. Also discussed in this paper is an
iterative importance sampling strategy, where the information in the
previous iteration is used to improve the importance function for the
transformed parameters. Smith et al. (\citeyear{smithetal87}) suggest using
``quasi-random'' sequences on the $k$-dimensional hypercube and reference
the preprint of Shaw (\citeyear{shaw88a}).

The paper by Smith et al. (\citeyear{smithetal87}) sets the computational theme of the
issue, as many of the other papers make use of either Gauss--Hermite
quadrature methodology or importance sampling. van Dijk, Hop and Louter
(\citeyear{vandijketal}) present the details of an algorithm for the computation of
posterior moments and densities based on importance sampling,
specifically that the importance sampling function can be adapted based
on the output of the previous iteration. In van Dijk, Hop and Louter
(\citeyear{vandijketal}), the posterior mean and the posterior covariance matrix based on
the output of the previous iteration is used to update the parameters
of the multivariate Student $t$ importance function.

Stewart (\citeyear{stewart}) illustrates how (nonadaptive) importance sampling can be
used in the context of hierarchical Bayesian models. Grieve (\citeyear{grieve87}),
Marriott (\citeyear{marriott87}), Naylor (\citeyear{naylor}) and Shaw (\citeyear{shaw87}) overcome the analytic
intractability of the posterior using the adaptive Gauss--Hermite
integration strategy of Naylor and Smith (\citeyear{naylorsmith}). Van Der Merwe and
Groenewald (\citeyear{vander}) approximate the posterior distribution with a Pearson
distribution, while Achcar, Bolfarine and Pericchi~(\citeyear{achcaretal}) make use of
the Laplace approximation discussed in Tierney and Kadane\break (\citeyear{tierneykadane}).
Spiegelhalter (\citeyear{spiegelhalter}), in his treatment of evidence propagation in
expert systems, briefly mentions stochastic relaxation (Geman, \citeyear{geman88b})
as a possible technique but does not use it in his
treatment.\footnote{See also Pearl (\citeyear{pearl}).} Finally, O'Hagan (\citeyear{ohagan})
expressed strong objection to using Monte Carlo methods in Bayesian
inference---even going as far as giving his paper the title ``Monte
Carlo is fundamentally unsound''.

\subsection{The Third Valencia International Meeting: June~1--5, 1987}

The Third Valencia International Meeting on Baye\-sian Statistics was
held on June 1--5, 1987. According to the Preface of the Proceedings
(see Bernardo et al., \citeyear{bernardoetal}), the scientific program consisted of 31
invited papers, each with discussion, and 33 refereed contributed
papers: ``The selection of topics, authors and discussants ensures that
those Proceedings provide a definitive up-to-date overview of current
concerns and activity in \textit{Bayesian Statistics}, encompassing a wide range
of theoretical and applied research.''

As was seen in the paper by Smith et al. (\citeyear{smithetal85}), as well as the special
issue of \textit{The Statistician}, the key computational approaches
vying for contention in this Proceedings are approximations based on
Laplace's method, Gaussian quadrature numerical integration---possibly
implemented in an iterative manner, and importance sampling---possibly
embellished with an adaptive procedure to update the importance
function. A survey of available software for Bayesian analysis was
presented by Goel (\citeyear{goel}) and this paper was discussed by van Dijk
(\citeyear{vandijk}). In the body of the discussion, van Dijk (\citeyear{vandijk}) proposes a
decision tree to determine which computational technique is most
suitable for the problem at hand. He notes that if the posterior is
``reasonably well behaved in the sense that it is unimodal, continuous,
proper, not too skewed,'' then a Laplace approximation approach or
possibly an importance sampling approach using a Student $t$ importance
function, as discussed in van Dijk, Hop and Louter~(\citeyear{vandijketal}), should be
used. If a transformation of the parameters results in a ``more
regular shape of the posterior,'' then he suggests Naylor and Smith~(\citeyear{naylorsmith}). If the posterior distribution is unimodal, but not much more
is known, then van Dijk suggests importance sampling with adaptive
importance functions. In this regard, van Dijk (\citeyear{vandijk}) references the
preprint of Geweke (\citeyear{geweke}) who also advocates for importance sampling
and who remarks, ``Integration by Monte Carlo is an attractive research
tool because it makes numerical problems much more routine than do
other numerical integration methods.''

A second paper on Bayesian software was presented by Smith (\citeyear{smith88}). In
the first sentence of the paper's abstract, he notes, ``Recent
developments in methods of numerical integration and approximation, in
conjunction with hardware trends towards the widespread availability of
single-user workstations which combine floating-point arithmetic power
with sophisticated graphics facilities in an integrated interactive
environment, would seem to have removed whatever excuses were hitherto
historically available for the lack of any generally available form of
Bayesian software.'' Smith argues that given advances in algorithm
development, as well as the movement from large mainframes to
workstations, the time is ripe for the development of Bayesian
software. Smith points out on page 433 that the computational tools he
has in mind are approximations based on asymptotic expansions,
numerical integration methods based on quadrature and importance
sampling methodology. In discussing the state of the field, Zellner
(\citeyear{zellner}) notes, ``\ldots\ it is concluded that a Bayesian era in econometrics
and statistics has emerged.''

DuMouchel (\citeyear{dumouchel}), Albert (\citeyear{albert}),
Poirier (\citeyear{poirier}) and Sweeting (\citeyear{sweeting})
employ or suggest the use of the Laplace approximation referencing
Tierney and Kadane (\citeyear{tierneykadane}) or related references.
Morris (\citeyear{morris88}) proposes
an approach based on the Pearson family which can be used to generalize
the method of Laplace. Kass, Tierney and Kadane (\citeyear{kasstierneykadane}) discuss the
Laplace approximation in detail, extending the theory, providing
examples and discussing implementation in the \texttt{S} computing
environment.

Kim and Schervish (\citeyear{kimschervish}) make extensive use of the Gauss--Hermite
approach as presented in Smith et al.~(\citeyear{smithetal85}) and the discussants of
this paper suggest the use of spherical integration rules (as
implemented in \textit{Bayes 4}). Shaw (\citeyear{shaw88b}) discusses several
approaches to numerical integration, with emphasis on Gauss--Hermite
and importance sampling with quasi-random sequences (see Shaw,
\citeyear{shaw87}, \citeyear{shaw88a}). Grieve~(\citeyear{grieve88}) uses Gauss--Hermite in the analysis
of $\mathit{LD}50$ experiments, Marriott (\citeyear{marriott88}) uses these methods (referencing
\textit{Bayes~4}) in the context of ARMA time series models and Pole
(\citeyear{pole}) in the context of state-space models. Schnatter (\citeyear{schnatter}) uses
generalized Laguerre integration for forecasting $\operatorname{AR}(p)$ time series
models.

Rubin (\citeyear{rubin88}) presents an overview\footnote{The presentation of this
algorithm comprised the bulk of his discussion of Tanner and Wong
(\citeyear{tannerwong}).} of his importance sampling based algorithm $\mathit{SIR}$, which he
proposes as a general approach for posterior simulation, distinguishing
it from the iterative MCMC approach in Tanner and Wong (\citeyear{tannerwong}), noting,
``The SIR (Sampling/Importance Resampling) algorithm is an ubiquitously
applicable noniterative algorithm for obtaining draws from an awkward
distribution: $M$ draws from an initial approximation are made, and
then $m < M$ draws are made from these with probability approximately
proportional to their importance ratios.''

A noted exception to the Laplace/numerical integration/importance
sampling approach to Bayesian computing is the paper by Geman (\citeyear{geman88a}).
In Section 2.5 on computing, Geman very clearly points out the basic
idea behind MCMC methods, ``Dynamics are simulated by producing a
Markov chain, $X(1), X(2),\ldots$ with transition
probabilities chosen so that \mbox{the equilibrium} distribution is the
posterior (Gibbs) distribution (2.4). One way to do this is with the
Metropolis algorithm (Metropolis et al., \citeyear{metropolisetal}). More convenient for
image processing is a variation we call \textit{stochastic
relaxation}.'' He then proceeds to present what we now refer to as the
full conditionals for the Gibbs sampler in the context of his problem.
However, neither in the discussion nor in the response is there
evidence to suggest that anyone at the time envisioned the MCMC
methodology presented in Geman~(\citeyear{geman88a}) having a broader impact to
general parametric statistical problems. If anything, the methodology
seems pigeon-holed as techniques for image analysis. Examining the
Preface of the Proceedings, we find that the paper is summarized as
``The important area of image-processing is reviewed by Geman.'' In
fact, Geman appeared to have doubts about the practical utility of MCMC
methods in his remark on page 169: ``On the other hand, it is indeed
difficult to find approaches that are as computationally expensive as
ours. In this regard, we view Monte Carlo optimization techniques as
\textit{research tools}. They are poor substitutes for the efficient
dedicated algorithms that should be developed when facing applications
involving a flow of data and a need for speedy analysis.'' Added to
this is his comment found on page 171 when referring to the algorithm
in Besag (\citeyear{besag}): ``Furthermore, this \textit{deterministic} algorithm
is typically far more efficient than stochastic relaxation methods.''

\section*{Acknowledgments}
The authors thank the Associate Editor and four referees for their
thoughtful comments which greatly improved the paper. The authors also
thank Jun Liu for his comments and suggestions. The work of W.H.W. was supported by NSF.


\begin{thebibliography}{71}

\bibitem[\protect\citeauthoryear{}{1987}]{achcaretal}
\textsc{Achcar, J. A., Bolfarine, H.} and \textsc{Pericchi, L. R.}
(1987). {Transformation of survival data to an extreme value
distribution}. \textit{J. Roy. Statist. Soc. Ser. D Statistician}
\textbf{36} 229--234.

\bibitem[\protect\citeauthoryear{}{1988}]{albert}
\textsc{Albert, J.} (1988).
Bayesian estimation of Poisson means using
a hierarchical log-linear model. In \textit{Bayesian Statistics 3}
(J. M. Bernardo, M. H. Degroot,
D. V. Lindley and A. F. M. Smith, eds.) 519--531. Oxford Univ. Press, Oxford.
\MR{1008064}

\bibitem[\protect\citeauthoryear{}{1988}]{bernardoetal}
\textsc{Bernardo, J. M., Degroot, M. H., Lindley, D. V.} and \textsc
{Smith, A. F. M.}, \textsc{eds}. (1988).
\textit{Bayesian Statistics 3}. Oxford Univ. Press, Oxford.
\MR{1008039}

\bibitem[\protect\citeauthoryear{}{1986}]{besag}
\textsc{Besag, J.} (1986).
On the statistical analysis of dirty pictures.
\textit{J.~R. Stat. Soc. Ser. B Stat. Methodol.} \textbf{48} 259--302.
\MR{0876840}

\bibitem[\protect\citeauthoryear{}{1978}]{binder}
\textsc{Binder, K.} (1978).
\textit{Monte Carlo Methods in Statistical Physics}. Springer,
New York.
\MR{0555878}

\bibitem[\protect\citeauthoryear{}{1977}]{dempsteretal}
\textsc{Dempster, A. P., Laird, N. M.} and \textsc{Rubin, D. B.} (1977).
Maximum likelihood from incomplete data via the EM algorithm (with
discussion).
\textit{J. R. Stat. Soc. Ser. B Stat. Methodol.} \textbf{39} 1--38.
\MR{0501537}

\bibitem[\protect\citeauthoryear{}{1988}]{dumouchel}
\textsc{DuMouchel, W.} (1988).
A Bayesian model and a graphical elicitation procedure for multiple
comparisons. In \textit{Bayesian Statistics 3}
(J. M. Bernardo, M. H. Degroot, D. V. Lindley and
A. F. M. Smith, eds.) 127--146. Oxford Univ. Press, Oxford.
\MR{1008039}

\bibitem[\protect\citeauthoryear{}{1979}]{efron}
\textsc{Efron, B.} (1979).
Bootstrap methods: Another look at the Jackknife.
\textit{Ann. Statist.} \textbf{7} 1--26.
\MR{0515681}

\bibitem[\protect\citeauthoryear{}{1990}]{gelfandetal}
\textsc{Gelfand, A. E., Hills, S. E., Racine-Poon, A.} and \textsc
{Smith, A. F. M.} (1990).
Illustration of Bayesian inference in normal data models using Gibbs
sampling.
\textit{J. Amer. Statist. Assoc.} \textbf{85} 972--985.

\bibitem[\protect\citeauthoryear{}{1990}]{gelfandsmith}
\textsc{Gelfand, A. E.} and \textsc{Smith, A. F. M.} (1990).
Sampling-based approaches to calculating marginal densities.
\textit{J. Amer. Statist. Assoc.} \textbf{85} 398--409.
\MR{1141740}

\bibitem[\protect\citeauthoryear{}{1990}]{gelmanking}
\textsc{Gelman, A.} and \textsc{King, G.} (1990).
Estimating the electoral consequences of legislative redistricting.
\textit{J. Amer. Statist. Assoc.} \textbf{85} 274--282.

\bibitem[\protect\citeauthoryear{}{1988a}]{geman88a}
\textsc{Geman, S.} (1988a).
Experiments in Bayesian image analysis. In \textit{Bayesian
Statistics 3} (J. M. Bernardo, M. H. Degroot,
D. V. Lindley and A. F. M. Smith, eds.) 159--171. Oxford Univ. Press, Oxford.

\bibitem[\protect\citeauthoryear{}{1988b}]{geman88b}
\textsc{Geman, S.} (1988b).
Stochastic relaxation methods for image restoration and expert
systems. In \textit{Maximum
Entropy and Bayesian Methods in Science and Engineering (Vol. 2)}
(G. J. Erickson and C. R. Smith, eds.).
Kluwer, New York.
\MR{0970827}

\bibitem[\protect\citeauthoryear{}{1984}]{gemangeman}
\textsc{Geman, S.} and \textsc{Geman, D.} (1984).
Stochastic relaxation, Gibbs distributions and the Bayesian
restoration of images. \textit{IEEE Trans. Pattern Anal.
Mach. Intell.} \textbf{6} 721--741.

\bibitem[\protect\citeauthoryear{}{1989}]{geweke}
\textsc{Geweke, J.} (1989).
Bayesian inference in econometric models using Monte Carlo
integration. \textit{Econometrica} \textbf{57} 1317--1339.
\MR{1035115}

\bibitem[\protect\citeauthoryear{}{1991}]{geyer91}
\textsc{Geyer, C. J.} (1991).
Markov chain Monte Carlo maximum likelihood. In
\textit{Computing Science and Statistics: Proceedings of the 23rd
Symposium on the Interface} (E. Keramidas, ed.) 156--163. Interface Foundation,
Fairfax Station.

\bibitem[\protect\citeauthoryear{}{1995}]{geyer95}
\textsc{Geyer, C. J.} (1995).
Conditioning in Markov chain Monte Carlo.
\textit{J. Comput. Graph. Statist.} \textbf{4} 148--154.
\MR{1341319}

\bibitem[\protect\citeauthoryear{}{1988}]{goel}
\textsc{Goel, P. K.} (1988).
Software for Bayesian analysis: Current status and additional need.
In \textit{Bayesian Statistics 3} (J. M. Bernardo, M. H. Degroot, D. V. Lindley
and A. F. M. Smith, eds.) 173--188. Oxford
Univ. Press, Oxford.

\bibitem[\protect\citeauthoryear{}{1987}]{grieve87}
\textsc{Grieve, A. P.} (1987).
Applications of Bayesian software: Two examples.
\textit{J. Roy. Statist. Soc. Ser. D Statistician} \textbf{36} 283--288.

\bibitem[\protect\citeauthoryear{}{1988}]{grieve88}
\textsc{Grieve, A. P.} (1988).
A Bayesian approach to the analysis of LD50 experiments. In \textit
{Bayesian Statistics 3}
(J. M. Bernardo, M.~H. Degroot, D. V. Lindley and A. F. M. Smith, eds.)
617--630. Oxford Univ. Press, Oxford.

\bibitem[\protect\citeauthoryear{}{2003}]{gubernatis}
\textsc{Gubernatis, J. E.}, ed. (2003).
\textit{The Monte Carlo Method in the Physical Sciences: Celebrating
the 50th Anniversary of the Metropolis Algorithm}. Amer. Inst.
Phys., New York.

\bibitem[\protect\citeauthoryear{}{1964}]{hammersleyhandscomb}
\textsc{Hammersley, J. M.} and \textsc{Handscomb, D. C.} (1964).
\textit{Monte Carlo Methods}, 2nd ed. Chapman and Hall,
London.
\MR{0223065}

\bibitem[\protect\citeauthoryear{}{1970}]{hastings}
\textsc{Hastings, W. K.} (1970).
Monte Carlo sampling methods using Markov chains and their
applications. \textit{Biometrika} \textbf{57} 97--109.

\bibitem[\protect\citeauthoryear{}{2003}]{hitchcock}
\textsc{Hitchcock, D. B.} (2003).
A history of the Metropolis--Hastings algorithm. \textit{Amer.
Statist.} \textbf{57} 254--257.
\MR{2037852}

\bibitem[\protect\citeauthoryear{}{1996}]{huku}
\textsc{Hukushima, K}. and \textsc{Nemoto, K.} (1996).
Exchange Monte Carlo method and application to spin glass
simulations.
\textit{J. Phys. Soc. Japan} \textbf{65} 1604--1608.

\bibitem[\protect\citeauthoryear{}{1975}]{karlintaylor}
\textsc{Karlin, S.} and \textsc{Taylor, H. M.} (1975).
\textit{A First Course in Stochastic Processes}, 2nd ed. Academic
Press, New York.
\MR{0356197}

\bibitem[\protect\citeauthoryear{}{1997}]{kass}
\textsc{Kass, R. E.} (1997).
Review of ``Markov chain Monte Carlo in practice.''
\textit{J. Amer. Statist. Assoc.} \textbf{92} 1645--1646.

\bibitem[\protect\citeauthoryear{}{1988}]{kasstierneykadane}
\textsc{Kass, R. E., Tierney, L.} and \textsc{Kadane, J. B.} (1988).
Asymptotics in Bayesian computation. In \textit{Bayesian
Statistics 3} (J. M. Bernardo, M. H. Degroot,
D. V. Lindley and A. F. M. Smith, eds.) 261--278. Oxford Univ. Press, Oxford.
\MR{1008051}

\bibitem[\protect\citeauthoryear{}{1988}]{kimschervish}
\textsc{Kim, C. E.} and \textsc{Schervish, M. J.} (1988).
Stochastic models of incarceration careers. In \textit{Bayesian
Statistics 3}
(J. M. Bernardo, M. H. Degroot, D. V. Lindley and A. F. M. Smith, eds.) 279--305.
Oxford Univ. Press, Oxford.
\MR{1008052}

\bibitem[\protect\citeauthoryear{}{1978}]{kloek1}
\textsc{Kloek, T.} and \textsc{van Dijk, H. K.} (1978).
Bayesian estimates of equation system parameters: An application of
integration by Monte Carlo. \textit{Econometrica} \textbf{46} 1--19.

\bibitem[\protect\citeauthoryear{}{1980}]{kloek2}
\textsc{Kloek, T.} and \textsc{van Dijk, H. K.} (1980).
Further experience in Bayesian analysis using Monte Carlo
integration. \textit{J. Econometrics} \textbf{14} 307--328.

\bibitem[\protect\citeauthoryear{}{1988}]{li8}
\textsc{Li, K. H.} (1988).
Imputation using Markov chains.
\textit{J. Statist. Comput. Simul.} \textbf{30} 57--79.
\MR{1005883}

\bibitem[\protect\citeauthoryear{}{1998}]{chanhai}
\textsc{Liu, C., Rubin, D. B.} and \textsc{Wu, Y. N.} (1998).
Parameter expansion to accelerate EM: The PX-EM algorith.
\textit{Biometrika} \textbf{85} 755--770.
\MR{1666758}

\bibitem[\protect\citeauthoryear{}{1994}]{liuetal}
\textsc{Liu, J. S.}, \textsc{Wong, W. H.} and \textsc{Kong, A.} (1994).
Covariance structure of the Gibbs sampler with applications to the
comparisons of estimators and augmentation schemes.
\textit{Biometrika} \textbf{81} 27--40.
\MR{1279653}

\bibitem[\protect\citeauthoryear{}{1999}]{liu}
\textsc{Liu, J. S.} and \textsc{Wu, Y. N.} (1999).
Parameter expansion scheme for data augmentation.
\textit{J. Amer. Statist. Assoc.} \textbf{94} 1264--1274.
\MR{1731488}

\bibitem[\protect\citeauthoryear{}{2009}]{lunnetal}
\textsc{Lunn, D., Spiegelhalter, D. J., Thomas, A.} and \textsc{Best,
N.} (2009).
The BUGS project: Evolution, critique and future directions.
\textit{Stat. Med.} \textbf{28} 3049--3067.

\bibitem[\protect\citeauthoryear{}{1987}]{marriott87}
\textsc{Marriott, J.} (1987).
Bayesian numerical and graphical methods for Box--Jenkins time series.
\textit{J. Roy. Statist. Soc. Ser. D
Statistician} \textbf{36} 265--268.

\bibitem[\protect\citeauthoryear{}{1988}]{marriott88}
\textsc{Marriott, J.} (1988).
Reparametrization for Bayesian inference in ARMA time series. In
\textit{Bayesian Statistics 3}
(J. M. Bernardo, M. H. Degroot, D. V. Lindley and A. F. M. Smith,
eds.) 701--704. Oxford Univ. Press, Oxford.

\bibitem[\protect\citeauthoryear{}{1953}]{metropolisetal}
\textsc{Metropolis, N., Rosenbluth,
A. W., Rosenbluth, M. N., Teller, A. H.} and \textsc{Teller, E.} (1953).
Equation of state calculations by fast computing machines.
\textit{J. Chem. Phys.} \textbf{21} 1087--1092.

\bibitem[\protect\citeauthoryear{}{1987}]{morris87}
\textsc{Morris, C. N.} (1987).
Comment on ``The calculation of posterior distributions by data
augmentation'' by M. A. Tanner and W. H. Wong.
\textit{J. Amer. Statist. Assoc.} \textbf{82} 542--543.
\MR{0898357}

\bibitem[\protect\citeauthoryear{}{1988}]{morris88}
\textsc{Morris, C. N.} (1988).
Approximating posterior distributions and posterior moments. In
\textit{Bayesian Statistics 3}
(J. M. Bernardo, M. H. Degroot, D. V. Lindley and A. F. M. Smith,
eds.) 327--344. Oxford
Univ. Press, Oxford.
\MR{1008054}

\bibitem[\protect\citeauthoryear{}{1987}]{naylor}
\textsc{Naylor, J. C.} (1987).
Bayesian alternatives to $t$-tests.
\textit{J. Roy. Statist. Soc. Ser. D Statistician} \textbf{36} 241--246.

\bibitem[\protect\citeauthoryear{}{1982}]{naylorsmith}
\textsc{Naylor, J. C.} and \textsc{Smith, A. F. M.} (1982).
Applications of a method for the efficient computation of posterior
distributions.
\textit{J. Roy. Statist. Soc. Ser. C Appl.
Statist.} \textbf{31} 214--225.
\MR{0694917}

\bibitem[\protect\citeauthoryear{}{1987}]{ohagan}
\textsc{O'Hagan, A.} (1987).
Monte Carlo is fundamentally unsound.
\textit{J.~Roy.
Statist. Soc. Ser. D Statistician} \textbf{36} 247--249.

\bibitem[\protect\citeauthoryear{}{1987}]{pearl}
\textsc{Pearl, J.} (1987).
Evidential reasoning using stochastic simulation of causal models.
\textit{Artif. Intell.} \textbf{32} 245--257.
\MR{0885357}

\bibitem[\protect\citeauthoryear{}{1988}]{poirier}
\textsc{Poirier, D. J}. (1988).
Bayesian diagnostic testing in the general linear normal regression
model. In \textit{Bayesian Statistics 3}
(J. M. Bernardo, M. H. Degroot, D. V. Lindley and A. F. M. Smith,
eds.) 725--732. Oxford Univ. Press, Oxford.
\MR{1008083}

\bibitem[\protect\citeauthoryear{}{1988}]{pole}
\textsc{Pole, A.} (1988).
Transfer response models: a numerical approach. In \textit{Bayesian
Statistics 3}
(J. M. Bernardo,
M. H. Degroot, D. V. Lindley and A. F. M. Smith, eds.)
733--745. Oxford Univ. Press,
Oxford.

\bibitem[\protect\citeauthoryear{}{1987}]{ripley87}
\textsc{Ripley, B. D.} (1987).
\textit{Stochastic Simulation}. Wiley, New York.
\MR{0875224}

\bibitem[\protect\citeauthoryear{}{2010}]{robertcasella10}
\textsc{Robert, C.} and \textsc{Casella, G.} (2010).
A short history of Markov chain Monte Carlo---subjective recollections
from incomplete data.
In \textit{Handbook on Markov Chain Monte Carlo}.
Chapman and Hall/CRC Press, Boca Raton, FL.

\bibitem[\protect\citeauthoryear{}{1988}]{rubin88}
\textsc{Rubin, D. B.} (1988).
Using the SIR algorithm to simulate posterior distributions. In
\textit{Bayesian Statistics 3}
(J. M. Bernardo, M.~H. Degroot, D. V. Lindley and A. F. M. Smith,
eds.) 395--402. Oxford Univ. Press, Oxford.

\bibitem[\protect\citeauthoryear{}{1981}]{rubinstein}
\textsc{Rubinstein, R. Y.} (1981).
\textit{Simulation and the Monte Carlo Method}, 1st ed.
Wiley, New York.
\MR{0624270}

\bibitem[\protect\citeauthoryear{}{1988}]{schnatter}
\textsc{Schnatter, S.} (1988).
Bayesian forecasting of time series by Gaussian sum approximation. In
\textit{Bayesian Statistics 3}
(J. M. Bernardo, M. H. Degroot, D. V. Lindley and A. F. M. Smith,
eds.) 757--764. Oxford
Univ. Press, Oxford.
\MR{1008084}

\bibitem[\protect\citeauthoryear{}{1987}]{shaw87}
\textsc{Shaw, J. E. H.} (1987).
Numerical Bayesian analysis of some flexible regression models.
\textit{J. Roy. Statist. Soc. Ser. D
Statistician} \textbf{36} 147--153.

\bibitem[\protect\citeauthoryear{}{1988a}]{shaw88a}
\textsc{Shaw, J. E. H.} (1988a).
A quasirandom approach to integration in Bayesian statistics.
\textit{Ann. Statist.} \textbf{16} 895--914.
\MR{0947584}

\bibitem[\protect\citeauthoryear{}{1988b}]{shaw88b}
\textsc{Shaw, J. E. H.} (1988b).
Aspects of numerical integration and summarisation. In \textit{Bayesian
Statistics 3}
(J. M. Bernardo, M. H. Degroot, D. V. Lindley and A. F. M. Smith, eds.)
411--428. Oxford Univ. Press,
Oxford.
\MR{1008059}

\bibitem[\protect\citeauthoryear{}{1988}]{smith88}
\textsc{Smith, A. F. M.} (1988).
What should be Bayesian about Bayesian software? In \textit{Bayesian
Statistics 3}
(J. M. Bernardo, M. H. Degroot, D. V. Lindley and A. F. M. Smith, eds.)
429--435. Oxford Univ. Press,
Oxford.

\bibitem[\protect\citeauthoryear{}{1991}]{smith91}
\textsc{Smith, A. F. M.} (1991).
Bayesian computational methods.
\textit{Philos. Trans.
Roy. Soc. Lond. Ser. A} \textbf{337} 369--386.
\MR{1143728}

\bibitem[\protect\citeauthoryear{}{1987}]{smithetal87}
\textsc{Smith, A. F. M., Skene, A. M., Shaw, J. E. H.} and \textsc
{Naylor, J.~C.} (1987).
Progress with numerical and graphical methods for practical Bayesian
statistics.
\textit{J. Roy. Statist. Soc. Ser.
D Statistician} \textbf{36} 75--82.

\bibitem[\protect\citeauthoryear{}{1985}]{smithetal85}
\textsc{Smith, A. F. M., Skene, A. M.,
Shaw, J. E. H., Naylor, J.~C.} and \textsc{Dransfield, M.} (1985).
The implementation of the Bayesian paradigm. \textit{Commun.
Stat. Theory Methods} \textbf{14} 1079--1102.
\MR{0797634}

\bibitem[\protect\citeauthoryear{}{1987}]{spiegelhalter}
\textsc{Spiegelhalter, D. J.} (1987).
Coherent evidence propagation in expert systems.
\textit{J. Roy. Statist. Soc. Ser. D Statistician} \textbf{36} 201--210.

\bibitem[\protect\citeauthoryear{}{1990}]{spiegellaur}
\textsc{Spiegelhalter, D. J.} and \textsc{Lauritzen, S. L.} (1990).
Sequential updating of conditional probabilities on directed graphical
structures. \textit{Networks} \textbf{20} 579--605.
\MR{1064738}

\bibitem[\protect\citeauthoryear{}{1987}]{stewart}
\textsc{Stewart, L.} (1987).
Hierarchical Bayesian analysis using Monte Carlo integration:
Computing posterior distributions when there are many possible models.
\textit{J. Roy. Statist. Soc. Ser. D
Statistician} \textbf{36} 211--219.

\bibitem[\protect\citeauthoryear{}{1988}]{sweeting}
\textsc{Sweeting, T. J.} (1988).
Approximate posterior distributions in censored regression models. In
\textit{Bayesian Statistics 3}
(J. M. Bernardo, M. H. Degroot, D. V. Lindley and A. F. M. Smith,
eds.) 791--799. Oxford
Univ. Press, Oxford.
\MR{1008087}\

\bibitem[\protect\citeauthoryear{}{1987}]{swendsenwang}
\textsc{Swendsen, R. H.} and \textsc{Wang, J. S.} (1987).
Nonuniversal critical dynamics in Monte Carlo simulations.
\textit{Phys. Rev. Lett.} \textbf{58} 86--88.

\bibitem[\protect\citeauthoryear{}{1987}]{tannerwong}
\textsc{Tanner, M. A.} and \textsc{Wong, W. H.} (1987).
The calculation of posterior distributions by data augmentation (with
discussion). \textit{J. Amer. Statist. Assoc.}
\textbf{82} 528--550.
\MR{0898357}

\bibitem[\protect\citeauthoryear{}{1986}]{tierneykadane}
\textsc{Tierney, L.} and \textsc{Kadane, J. B.} (1986).
Accurate approximations for posterior moments and marginal densities.
\textit{J. Amer. Statist. Assoc.} \textbf{81} 82--86.
\MR{0830567}

\bibitem[\protect\citeauthoryear{}{1987}]{vander}
\textsc{van der Merwe, A. J.} and \textsc{Groenewald, P. C. N.} (1987).
Bayes and empirical Bayes confidence intervals in applied research.
\textit{J. Roy. Statist. Soc. Ser. D
Statistician} \textbf{36} 171--179.

\bibitem[\protect\citeauthoryear{}{1988}]{vandijk}
\textsc{van Dijk, H. K.} (1988).
Discussion of Goel. In \textit{Bayesian Statistics~3}
(J. M. Bernardo, M. H. Degroot, D. V. Lindley
and A. F. M. Smith, eds.)
187--188. Oxford Univ. Press, Oxford.
\
\bibitem[\protect\citeauthoryear{}{1987}]{vandijketal}
\textsc{van Dijk, H. K., Hop, J. P.} and \textsc{Louter, A. S.} (1987).
An algorithm for the computation of posterior moments and densities
using simple importance sampling. \textit{J. Roy.
Statist. Soc. Ser. D Statistician} \textbf{36} 83--90.

\bibitem[\protect\citeauthoryear{}{2001}]{vandyk}
\textsc{van Dyk, D. A.} and \textsc{Meng, X. L.} (2001).
The art of data augmentation.
\textit{J. Comput.
Graph. Statist.} \textbf{10} 1--50.
\MR{1936358}

\bibitem[\protect\citeauthoryear{}{1988}]{zellner}
\textsc{Zellner, A. }(1988).
A Bayesian era. In \textit{Bayesian Statistics 3}
(J.~M. Bernardo, M. H. Degroot, D. V. Lindley and
A. F. M. Smith, eds.) 509--516.
Oxford Univ. Press, Oxford.
\vspace*{-2pt}
\MR{1008063}
\end{thebibliography}
\end{document}